\documentclass[conference]{IEEEtran}
\IEEEoverridecommandlockouts
\usepackage{subfigure}
\usepackage{cite}
\usepackage{amsmath,amssymb,amsfonts}
\usepackage{algorithmic}
\usepackage{graphicx}
\usepackage{textcomp}
\usepackage{xcolor}
\usepackage{url}
\begin{document}

\title{WVSC: Wireless Video Semantic Communication with Multi-frame Compensation}
\author{Bingyan Xie, Yongpeng Wu, Yuxuan Shi, Biqian Feng, Wenjun Zhang, Jihong Park, Tony Q.S. Quek
	
	\thanks{Bingyan Xie, Yongpeng Wu, Biqian Feng, and Wenjun Zhang are with the Department of Electronic Engineering, Shanghai Jiao Tong University, Shanghai 200240, China (e-mail:bingyanxie, yongpeng.wu, fengbiqian, zhangwenjun@sjtu.edu.cn).}
	\thanks{Yuxuan Shi is with the School of Cyber and Engineering, Shanghai Jiao Tong University, Shanghai 200240, China (e-mail:ge49fuy@sjtu.edu.cn).}
	\thanks{Jihong Park and Tony Q.S. Quek are with the ISTD Pillar, Singapore University of Technology of Design, 8 Somapah Rd, Singapore 487372 (e-mail:jihong\_park, tonyquek@sutd.edu.sg)}
}
\maketitle
\begin{abstract}
Existing wireless video transmission schemes directly conduct video coding in pixel level, while neglecting the inner semantics contained in videos. In this paper, we propose a wireless video semantic communication framework, abbreviated as WVSC, which integrates the idea of semantic communication into wireless video transmission scenarios. WVSC first encodes original video frames as semantic frames and then conducts video coding based on such compact representations, enabling the video coding in semantic level rather than pixel level. Moreover, to further reduce the communication overhead, a reference semantic frame is introduced to substitute motion vectors of each frame in common video coding methods. At the receiver, multi-frame compensation (MFC) is proposed to produce compensated current semantic frame with a multi-frame fusion attention module. With both the reference frame transmission and MFC, the bandwidth efficiency improves with satisfying video transmission performance. Experimental results verify the performance gain of WVSC over other DL-based methods (e.g. DVSC) about 1 dB and traditional standardized schemes (e.g. H.265+5G LDPC) about 0.7 dB in terms of PSNR.
\end{abstract}


\section{Introduction}

The development of mobile Internet has triggered the prosperity of different video services. Various video-related applications have emerged, e.g. Internet of things, virtual reality, and smart city, which contribute to the major Internet traffic nowadays. To support these applications with huge amount of video data, the efficient wireless video communication technology is of great demand.

When it comes to the wireless video transmission, the key challenge is how to exploit both intra-frame correlation and inter-frame correlation to efficiently compress original videos while ensuring accurate data transmission simultaneously. Wireless video transmission schemes commonly adopt standardized video codecs such as H.264 \cite{264} or H.265 \cite{265} for source coding, separately followed by the channel codec such as low density parity check (LDPC). These separated source-channel
coding (SSCC) schemes are obviously not optimal and would experience serious cliff effect. To address this, Deep learning (DL)-based schemes, especially semantic communications \cite{LCFSC,dvc,dvsc}, serve as promising solutions for innovating existing SSCC-based wireless video transmission frameworks. Existing works \cite{dvc,dvsc} have successfully conducted DL techniques into wireless video transmission designs. For example, Jiang et al. \cite{dvc} have proposed a wireless semantic video conferencing system to ensure the keypoint transmission in video conference. Niu et al. \cite{dvsc} have proposed an end-to-end DL-enabled video semantic communication system which conducts SNR-adaptive channel codec along with semantic repairment to combat multi-dimensional noise.

\begin{figure}[htbp]
	\centering
	\includegraphics[width=3.5in]{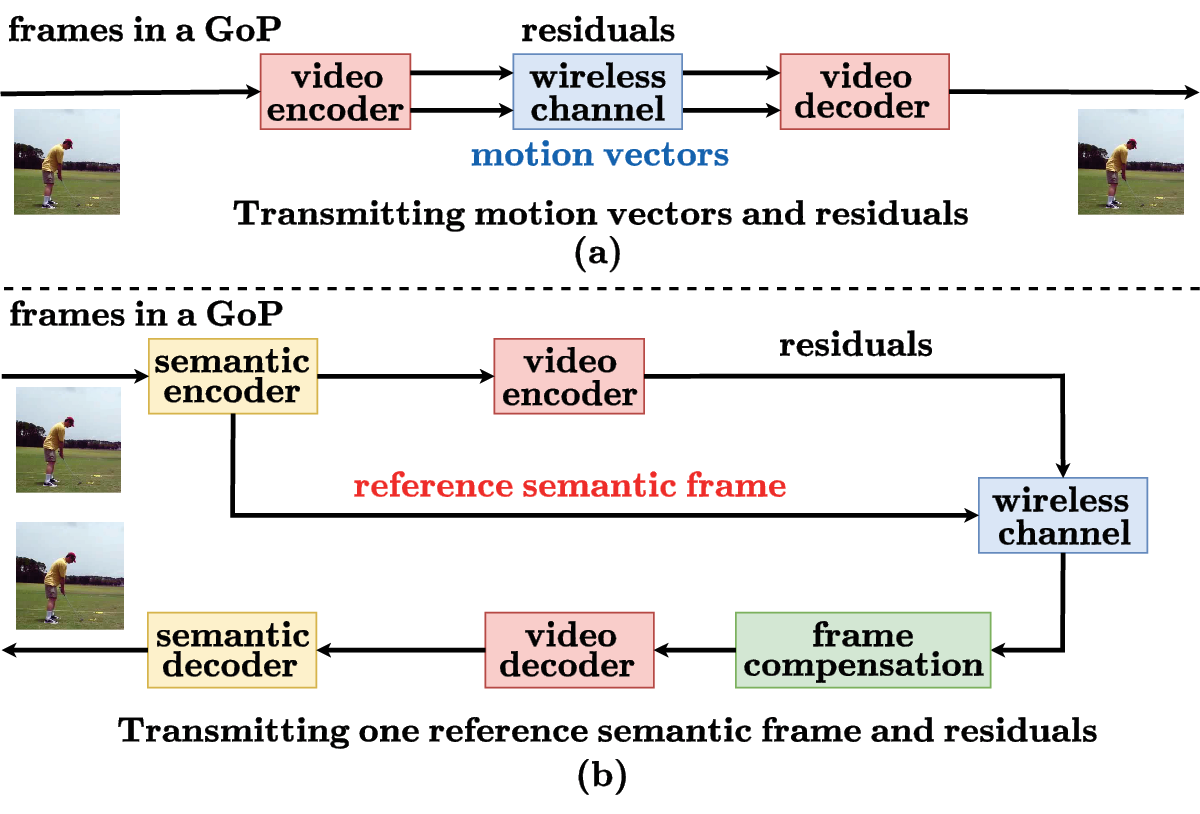}
	\caption{Different structures of wireless video transmission frameworks. (a) pixel-level wireless video transmission structure. (b) proposed semantic-level wireless video transmission structure.}
	\label{fig_1}
\end{figure}

Despite the success of above DL-based wireless video communication framework, e.g. \cite{dvsc}, such transmitted codewords are all related to the original pixel level, which neglects the semantic meaning intrinsically embedded in the data source, hindering its performance for efficient video transmission. As shown in Fig. \ref{fig_1}(a), pixel-level wireless video transmission structure conducts video coding in pixel level and transmits both motion vectors and residuals to the receiver, which brings extra communication overhead. To tackle this problem, inspired by the feature-level coding scheme \cite{fvc}, Fig. \ref{fig_1}(b) illustrates the proposed semantic-level wireless video transmission structure, integrating semantic communication and deep video coding together to improve the wireless video transmission performance. Different from existing video transmission frameworks \cite{dvsc} which are obliged to transmit both motion vectors and residuals of each frame in a group of pictures (GoPs), the proposed framework introduces the reference semantic frame to substitute motion vectors of each frame for transmission. The reference semantic frame refers to semantic I frame and other frames in the same GoP are semantic P frames. Thus, only specific semantic I frames along with residuals of semantic P frames are required to be transmitted, greatly reducing the communication redundancy. Meanwhile, \cite{dvc} utilizes the I frame as shared semantic knowledge and transmits only keypoints for P frame restoration. Such straightforward method is only applied for wireless video conferencing with simple facial changes, which would face serious performance degradation under diverse video contents. As such, to exploit the temporal relationship among consecutive frames, multi-frame compensation module is proposed at the receiver which utilizes previously reconstructed frames as extra motion information for combating channel noise and motion difference. The contributions of this paper are summarized as follows

\begin{figure*}[htbp]
	\centering
	\includegraphics[width=5.8in]{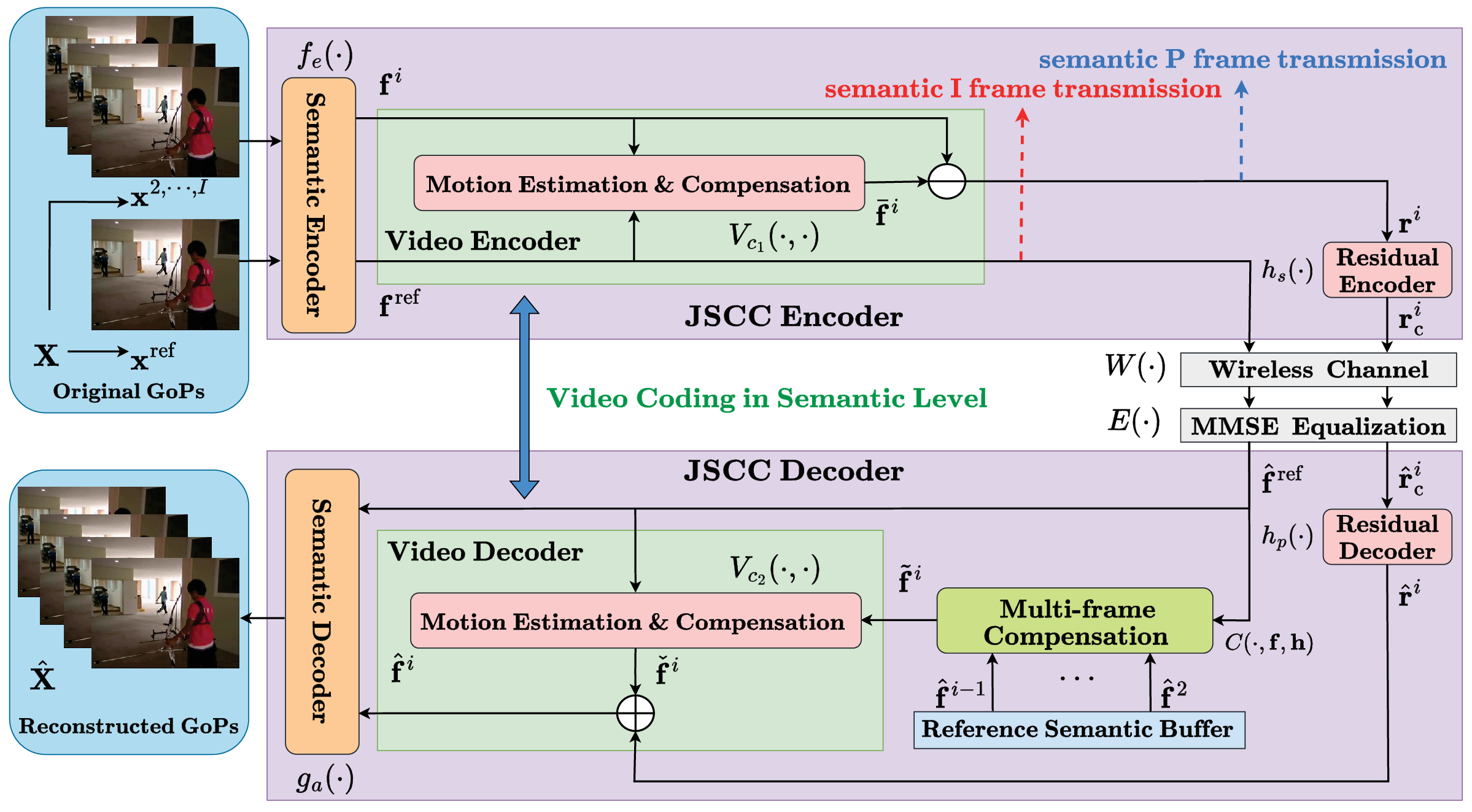}
	\caption{The proposed WVSC framework. The video is transmitted by a series of GoPs, which is divided into reference frame $\mathbf{x}^\mathrm{ref}$ and current frame $\mathbf{x}^i$. $\mathbf{x}^\mathrm{ref}$ is directly coded and transmitted through semantic coding while $\mathbf{x}^i$ is compressed by both semantic coding and video coding aided by $\mathbf{x}^\mathrm{ref}$.}
	\label{fig_2}
\end{figure*}
\begin{enumerate}
	\item{}
	To enhance bitrate saving by reducing redundant I and P frames, we propose a novel wireless video transmission framework, coined WVSC. The main idea of WVSC is to conduct DL-based video coding based on extracted semantics rather than directly on pure video signals. In this way, both the merits of semantic coding for alleviating intra-frame redundancy and video coding for inter-frame redundancy are utilized for more efficient wireless video transmission.
	\item{}
	To reduce the transmission redundancy in a single GoP, we introduce the reference frame for video coding-based semantic transmission. Instead of transmitting motion vectors for every semantic frame, the reference frame is transmitted only once in a GoP. In this way, only key semantic frames and residual information are transmitted, greatly saving the communication overhead.
	\item{}
	To reconstruct the semantic frames at the receiver, we propose a multi-frame compensation module to polish the received reference frame into the current frame as much as possible. Previously received frames are employed through a multi-frame fusion attention (MFA) module to reconstruct the current frame, thus providing refined semantic frames at the receiver.
\end{enumerate}

Notations: $\mathbb{R}$ and $\mathbb{C}$ refer to the real and complex number sets, respectively. $\mathcal{CN}\left (\mu, \sigma^2 \right)$ denotes a complex Gaussian distribution with mean $\mu$ and variance $\sigma^2$. $\mathrm{diag}(\cdot)$ refer to the diagonalization operations between a vector and its corresponding diagonal matrix. $|\cdot|$ refers to computing the modulus of a complex number. $\ast$ refers to the element-wise multiplication. $\mathbf{I}$ denotes the unit matrix. The operator $\left(\cdot\right)^{T}$ denotes the matrix transpose.

\section{System Model and Proposed Framework}
In this section, we describe the system model of WVSC framework by both semantic I frame transmission and semantic P frame transmission.

\subsection{Semantic I frame transmission}
We consider a GoP-based wireless video transmission system. Given a video sequence $\mathcal{X}= \left \{\mathbf{X}_1,\mathbf{X}_2,\cdots,\mathbf{X}_N \right\}$, where $\mathbf{X}_n\in\mathbb{R}^{I\times H\times W\times3}, n=1,\cdots, N$, which is composed of $N$ successive GoPs. For an arbitrary GoP $\mathbf{X} = \left\{\mathbf{x}^1,\mathbf{x}^2,\cdots,\mathbf{x}^I\right\}\in\mathcal{X}$, where $I$ is the total number of frames in a GoP, $\mathbf{x}^i\in \mathbb{R}^{H\times W\times3}$, we treat the first frame $\mathbf{x}^1$ as the reference frame $\mathbf{x}^{\mathrm{ref}}$. The reference frame, which is also the I frame in a GoP, is communicated using DL-based JSCC, i.e., the semantic encoder and decoder in Fig. \ref{fig_2}, similar to the simple image semantic communications. The semantic encoder, $f_e(\cdot): \mathbb{R}^{I\times H\times W\times 3}\longrightarrow \mathbb{R}^{I\times L}$, encodes $\mathcal{X}$ into the semantic sequence, $\mathcal{F}=\left\{\mathbf{f}^1,\mathbf{f}^2,\cdots,\mathbf{f}^I\right\}$, where $L$ is the code length for the extracted semantic frame with $\mathbf{f}^i\in \mathbb{R}^{L}$. Here $\mathbf{f}^{\mathrm{ref}}=\mathbf{f}^1$ is the semantic I frame. Then, the semantic I frame is transmitted through wireless channels and equalized by the minimum mean square error (MMSE) channel equalization. It can be formulated as
\begin{align}
	\hat{\mathbf{f}}^{\mathrm{ref}}=\mathbf{H}_s\ast\mathbf{f}^{\mathrm{ref}}+\mathbf{H}_n\ast\mathbf{n}, 
\end{align}
where $\mathbf{H}_s=\mathrm{diag}(|\mathbf{h}_d|^2(|\mathbf{h}_d|^2+\sigma^{2}\mathbf{I})^{-1})\in\mathbb{R}^{L}$ along with $\mathbf{H}_n=\mathrm{diag}(\mathbf{h}_d^T(|\mathbf{h}_d|^2+\sigma^{2}\mathbf{I})^{-1})\in\mathbb{C}^{L}$ refers to the channel equalization parameters, $\mathbf{h}_d = \mathrm{diag}(\mathbf{h})\in \mathbb{C}^{L\times L}$, $\mathbf{h}\in \mathbb{C}^{L}$ denotes the Rayleigh channel fading index following the distribution of $\mathcal{CN}(0,1)$, $\mathbf{n}\in \mathbb{C}^{L}$ is the complex Gaussian channel noise vector whose component has zero mean and covariance $\sigma^{2}$.

At the receiver, the semantic decoder, $g_a(\cdot): \mathbb{R}^{I\times L}\longrightarrow \mathbb{R}^{I\times H\times W\times 3}$, converts $\hat{\mathcal{F}}=\left\{\hat{\mathbf{f}^1},\hat{\mathbf{f}^2},\cdots,\hat{\mathbf{f}^I}\right\}$ into the final reconstructed video sequence $\hat{\mathcal{X}}=\left\{\hat{\mathbf{X}}^1,\hat{\mathbf{X}}^2,\cdots,\hat{\mathbf{X}}^I\right\}$. The whole transmission process for the semantic I frame can be formulated as
\begin{align}
	\mathbf{x}^{\mathrm{ref}} \xrightarrow{f_e(\cdot)} \mathbf{f}^{\mathrm{ref}} \xrightarrow{E\left(W(\cdot)\right)} \hat{\mathbf{f}}^{\mathrm{ref}} \xrightarrow{g_a(\cdot)} \mathbf{\hat{x}}^{\mathrm{ref}},
\end{align}
where $W(\cdot)$ implies wireless channels and $E(\cdot)$ refers to the MMSE channel equalization.

\subsection{Semantic P frame transmission}
For the semantic P frame transmission, non-reference semantics $\mathbf{f}^i$ with $i\geq 2$, are communicated using not only DL-based JSCC but also additional pre-/post-processing. They are processed by the motion estimation $\&$ compensation video encoder, $V_{c_1}(\cdot, \mathbf{f}^{\mathrm{ref}}): \mathbb{R}^{L}\longrightarrow \mathbb{R}^{L}$, along with the reference semantics $\mathbf{f}^{\mathrm{ref}}$ to acquire the predicted semantic frame $\bar{\mathbf{f}}^i\in \mathbb{R}^{L}$. Then, the residual $\mathbf{r}^i\in \mathbb{R}^{L}$ can be computed through the subtraction between $\mathbf{f}^i$ and $\bar{\mathbf{f}}^i\in \mathbb{R}^{L}$. In this way, the transmission for the P frame semantics are substituted by the corresponding sparse residuals which is much easier to be compressed than original video frames. Notably, in common video coding schemes, $\bar{\mathbf{f}}^i$ is known for both compression and decompression process. However, for wireless video transmission scenarios, $\bar{\mathbf{f}}^i$ is obviously unknown at the receiving end. Instead of transmitting motion vectors for every semantic P frame to recover $\bar{\mathbf{f}}^i$, we utilize the consistent reference semantic frame $\mathbf{f}^{\mathrm{ref}}$ as the substitute for the motion vectors, greatly saving the communication cost. After that, the residuals are further compressed by the residual encoder, $h_s(\cdot): \mathbb{R}^{L}\longrightarrow \mathbb{R}^{L_1}$, to provide adjustable transmission rates for the semantic I frame and compressed residuals $\mathbf{r}_\mathrm{c}^i$, commonly preallocating more transmission rate for the semantic I frame to ensure the latter deep video decoding performances at the receiver. In this way, the semantic P frame is converted to the compactly compressed residuals for transmission. For simplicity, the transmitted $\mathbf{r}_\mathrm{c}^i$ in a GoP share the same channel states with the semantic I frame. The wireless transmission process for the compressed residuals of semantic P frames can be formulated as
\begin{align}
	\hat{\mathbf{r}}_\mathrm{c}^i=\mathbf{H}_{sr}\ast\mathbf{r}_\mathrm{c}^i+\mathbf{H}_{nr}\ast\mathbf{n}_r, 
\end{align}
where $\mathbf{H}_{sr}=\mathrm{diag}(|\mathbf{h}_{dr}|^2(|\mathbf{h}_{dr}|^2+\sigma^{2}\mathbf{I})^{-1})\in\mathbb{R}^{L_1}$, $\mathbf{H}_{nr}=\mathrm{diag}(\mathbf{h}_{dr}^T(|\mathbf{h}_{dr}|^2+\sigma^{2}\mathbf{I})^{-1})\in\mathbb{C}^{L_1}$, $\mathbf{h}_{dr} = \mathrm{diag}(\mathbf{h}_r)\in \mathbb{C}^{L_1\times L_1}$, $\mathbf{h}_r\in \mathbb{C}^{L_1}$, and $\mathbf{n}_r\in \mathbb{C}^{L_1}$ are the corresponding transmission parameters for the residuals similar to Eq. (1).

At the receiver, the residual decoder, $h_p(\cdot): \mathbb{R}^{L_1}\longrightarrow \mathbb{R}^{L}$, recovers the residuals as $\hat{\mathbf{r}}^i \in \mathbb{R}^{L}$. With the received $\hat{\mathbf{f}}^{\mathrm{ref}}$ and $\hat{\mathbf{r}}^i$, we obtain the reconstructed semantic sequence, $\hat{\mathcal{F}}=\left\{\hat{\mathbf{f}}^1,\hat{\mathbf{f}}^2,\cdots,\hat{\mathbf{f}}^I\right\}$, using a series of frame compensation and video decoding operations. The multi-frame compensation module, $C(\cdot,\mathbf{f}, \mathbf{h}): \mathbb{R}^{L}\longrightarrow \mathbb{R}^{L}$, is employed to polish $\mathbf{\hat{f}}^{\mathrm{ref}}$ into the current semantic P frame $\mathbf{f}^i$ as much as possible, where $\mathbf{f}=\left\{\mathbf{\hat{f}}^{i-1},\cdots,\mathbf{\hat{f}}^{2}\right\}$ refers to the previously reconstructed frames. $\mathbf{\tilde{f}}^i \in\mathbb{R}^{L}$ refers to the polished current semantic frame after multi-frame compensation. After multi-frame compensation, through motion estimation $\&$ compensation video decoder $V_{c_2}(\cdot, \mathbf{\hat{f}}^{\mathrm{ref}}): \mathbb{R}^{L}\longrightarrow \mathbb{R}^{L}$, the reconstructed predicted semantic frame $\mathbf{\check{f}}^i\in\mathbb{R}^{L}$ can be acquired. With $\hat{\mathbf{r}}^i$ and $\mathbf{\check{f}}^i$, we are able to acquire the current reconstructed semantic frames $\mathbf{\hat{f}}^i$ by adding them together. Finally, the semantic decoder decodes $\mathbf{\hat{f}}^i$ into final reconstructed video frames $\mathbf{\hat{x}}^i$. The whole process can be formulated as
\begin{align}
	\mathbf{x}^i \xrightarrow{f_e(\cdot)} \mathbf{f}^i \xrightarrow{V_{c_1}(\cdot, \mathbf{f}^{\mathrm{ref}})} \mathbf{\bar{f}}^i \xrightarrow{-\mathbf{f}^{\mathrm{ref}}} \mathbf{r}^i\xrightarrow{h_s(\cdot)}\mathbf{r}_\mathrm{c}^i, i\ne1
\end{align}
\begin{align}
	\mathbf{\hat{f}}^{\mathrm{ref}} \xrightarrow{C(\cdot,\mathbf{f}, \mathbf{h})} \mathbf{\tilde{f}}^i \xrightarrow{V_{c_2}(\cdot, \mathbf{\hat{f}}^{\mathrm{ref}})} \mathbf{\check{f}}^i \xrightarrow{+h_p(\mathbf{\hat{r}}_\mathrm{c}^i)} \mathbf{\hat{f}}^i \xrightarrow{g_a(\cdot)} \mathbf{\hat{x}}^i,i\ne1
\end{align}

\section{Details of WVSC Framework}
In this section, we present the structure and detailed designs of each module in the WVSC framework.

\subsection{Deep Video Coding in Semantic Level}
\begin{figure}[htbp]
	\centering
	\includegraphics[width=3.3in]{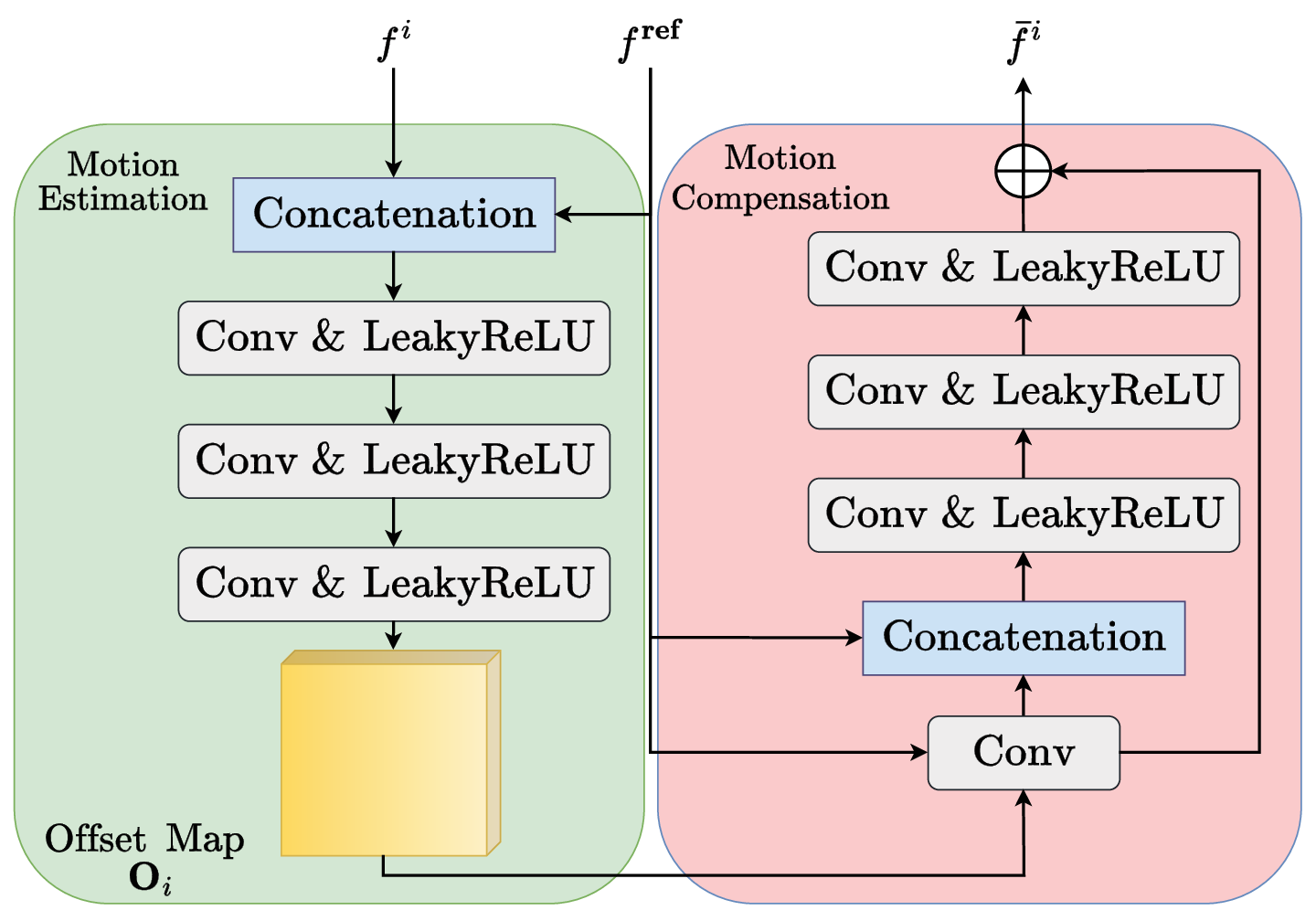}
	\caption{The structure of motion estimation $\&$ compensation network.}
	\label{fig_3}
\end{figure}
As illustrated in Section \uppercase\expandafter{\romannumeral2}, WVSC conducts deep video coding in semantic level. That is to say, the inter-frame correlation is exploited based on the semantics extracted by original frames. The structure of motion estimation $\&$ compensation video coder is shown in Fig. \ref{fig_3}. It is mentioned that the video encoder ${V_{c_1}(\cdot, \mathbf{f}^{\mathrm{ref}})}$ and decoder $V_{c_2}(\cdot, \mathbf{\hat{f}}^{\mathrm{ref}})$ share the same network structure. The video coder is composed of two parts: motion estimation and motion compensation. For the motion estimation, the current semantic frame $\mathbf{f}^i$ and the reference semantic frame $\mathbf{f}^{\mathrm{ref}}$ are concatenated together as input. Then through a lightweight structure composed of a series of convolutional layers and activation layers with LeakyReLU \cite{leakyrelu}, the motion offset map $O_i$ between reference and current semantics is generated. In other words, $O_i$ refers to the motion vector in common video coding frameworks. For the motion compensation, with the estimated offset map, the predicted semantic frame $\mathbf{\bar{f}}^i$ can be produced by the motion compensation network which is based on the residual convolutional networks.

\subsection{Multi-frame Compensation for the Reference Frame}
\begin{figure}[htbp]
	\centering
	\includegraphics[width=3.0in]{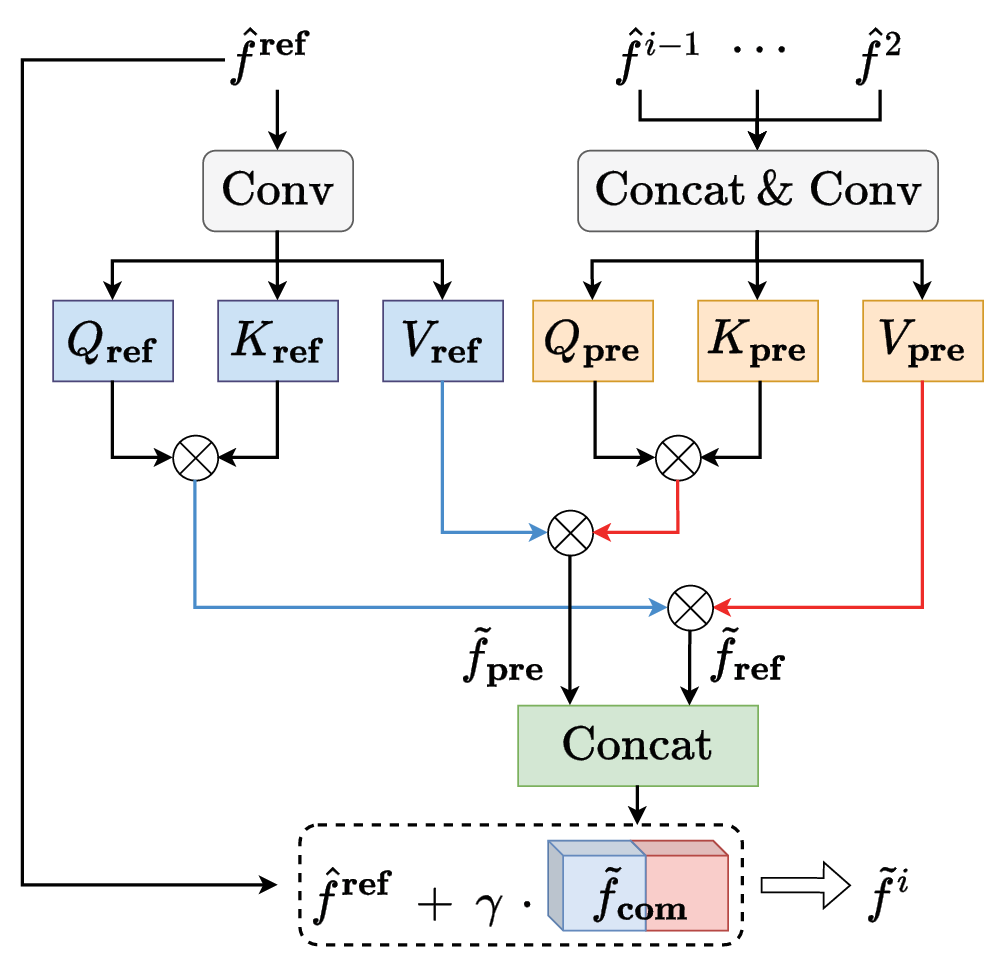}
	\caption{The architecture of MFA module.}
	\label{fig_4}
\end{figure}
For the received reference semantic frame, it is confronted with both the wireless channel noise and the inter-frame shift compared to the actual current semantic frame. In this way, we are obliged to compensate the reference semantic frame. Moreover, previously reconstructed semantic frames in the same GoP contain the semantic shift information and channel state information for the current frame transmission. As such, introducing previous semantic frames to compensate the reference semantic frame is a straightforward choice.

The detailed architecture of proposed MFA module is shown in Fig. \ref{fig_4}. The structure divides the computation process into two stages. One is the attention matrices generation. Two pairs of attention matrices $\mathbf{Q}_{\mathrm{ref}}$, $\mathbf{K}_{\mathrm{ref}}$, $\mathbf{V}_{\mathrm{ref}}$ and $\mathbf{Q}_{\mathrm{pre}}$, $\mathbf{K}_{\mathrm{pre}}$, $\mathbf{V}_{\mathrm{pre}}$ are generated based on the reference semantic frame $\mathbf{\hat{f}}^{\mathrm{ref}}$ and several previous semantic frames $\mathbf{\hat{f}}^{i-1},\cdots,\mathbf{\hat{f}}^{2}$, respectively. The other is the cross attention multiplication \cite{cross}. Through multiplication between crossed attention matrices, the previous semantics $\mathbf{\tilde{f}}_{\mathrm{pre}}$ and reference semantics $\mathbf{\tilde{f}}_{\mathrm{ref}}$ are computed. Finally, through residual connection, the compensated current semantic frame is presented. The whole process can be formulated as
\begin{align}
	\mathbf{\tilde{f}}_{\mathrm{ref}} & = \Phi(\mathbf{Q}_{\mathrm{ref}}\mathbf{K}_{\mathrm{ref}}^T)\mathbf{V}_{\mathrm{pre}},
\end{align}
\begin{align}
	\mathbf{\tilde{f}}_{\mathrm{pre}} & = \Phi(\mathbf{Q}_{\mathrm{pre}}\mathbf{K}_{\mathrm{pre}}^T)\mathbf{V}_{\mathrm{ref}},
\end{align}
\begin{align}
	\mathbf{\tilde{f}}_{\mathrm{com}} & = \mathrm{Con}(\mathbf{\tilde{f}}_{\mathrm{pre}}, \mathbf{\tilde{f}}_{\mathrm{ref}}),
\end{align}
\begin{align}
	\mathbf{\tilde{f}}^i & = \mathbf{\hat{f}}^{\mathrm{ref}} + \gamma \mathbf{\tilde{f}}_{\mathrm{com}},
\end{align}
where $\Phi(\cdot)$ is the softmax function, $\mathrm{Con}(\cdot,\cdot)$ refers to the concatenation between two feature matrices.

\subsection{Correlation between the I frame and the P frame}
For the WVSC, I frames and P frames have distinct transmission characteristics. Since the I frame serves as the key reference for the subsequent P frames in a GoP, more bandwidth resources should be allocated to the semantic I frame than to the semantic P frames. In this way, the residual encoder and decoder act as a controller to adjust the allocated bitrate between the semantic I frame and subsequent semantic P frames, thereby providing unequal error protection for the overall wireless video transmission.

The WVSC is trained using a successive training strategy, meaning it is optimized not only within a single frame but also across different GoPs to jointly train the I frames and P frames. The end-to-end video reconstruction loss is formulated as follows
\begin{align}
	L_\mathrm{R}=\frac{1}{N\times I}\sum_{n=1}^{N}\sum_{i=1}^{I}||\hat{\mathbf{x}}_n^i-{\mathbf{x}}_n^i||^2,
\end{align}
where mean square error (MSE) is set as the default loss, $\hat{\mathbf{x}}_n^i$ and ${\mathbf{x}}_n^i$ refer to the $i$-th reconstructed video frame and original frame in the $n$-th GoP, respectively.

With the above designs, video coding schemes can be properly integrated into the semantic communication framework.

\section{Numerical Results}
In this section, we present numerical results to evaluate the effectiveness of proposed WVSC for wireless video.

\subsection{Experimental Setups}
\subsubsection{Datasets}

For the wireless video semantic transmission, we quantify the performances of proposed WVSC versus other benchmarks over the UCF101 \cite{ucf} dataset, which is a widely-used dataset for deep video compression. The dataset is splited into about 5:1 ratio for training and testing, respectively. During model training, images are randomly cropped to 128$\times$128$\times$3.

\begin{figure*}[htbp]
	\centering  
	\subfigure[PSNR for the reconstructed videos.]{
		\includegraphics[width=0.32\linewidth]{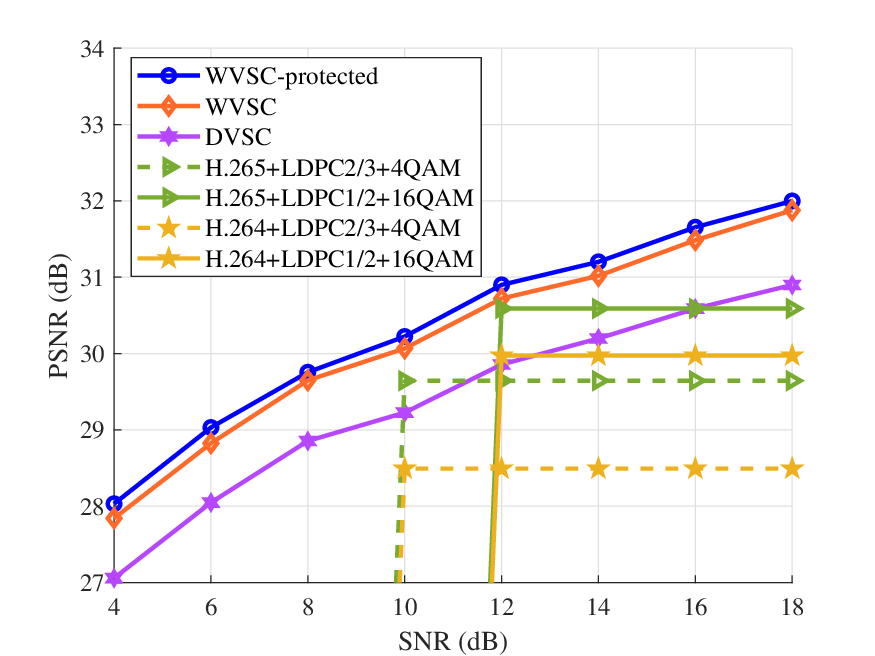}}
	\subfigure[MS-SSIM for the reconstructed videos.]{
		\includegraphics[width=0.32\linewidth]{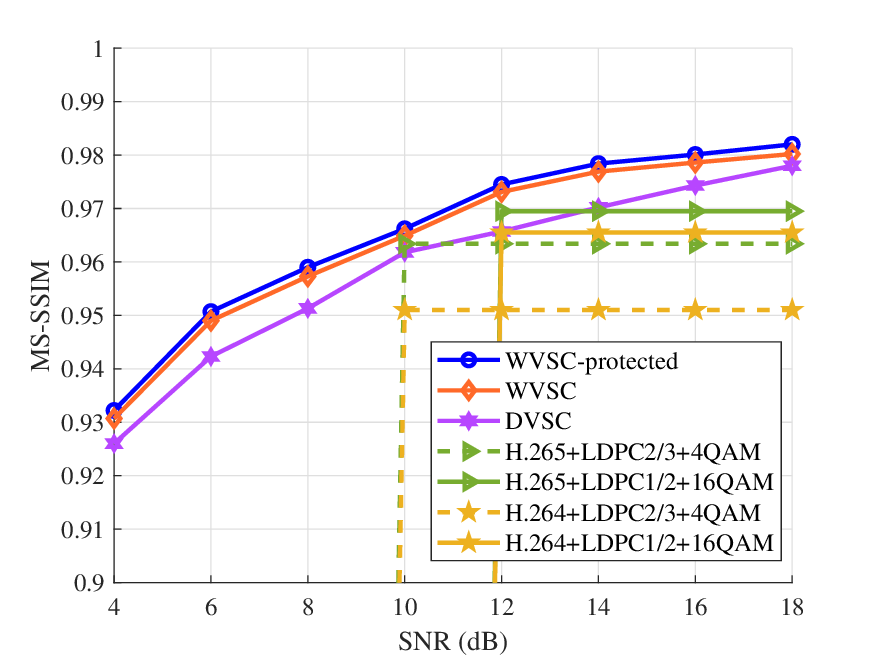}}
	\subfigure[LPIPS for the reconstructed videos.]{
		\includegraphics[width=0.32\linewidth]{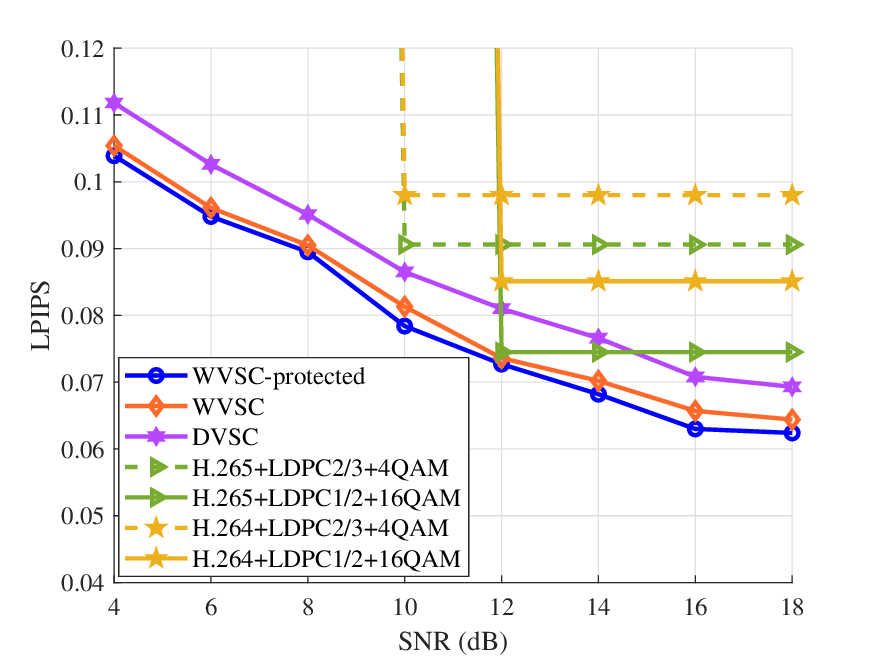}}
	\caption{Quality of the reconstructed images versus the SNRs under Rayleigh fading channels (CBR = 0.04).}
	\label{fig_5}
\end{figure*}
\begin{figure*}[htbp]
	\centering  
	\subfigure[PSNR for the reconstructed videos.]{
		\includegraphics[width=0.32\linewidth]{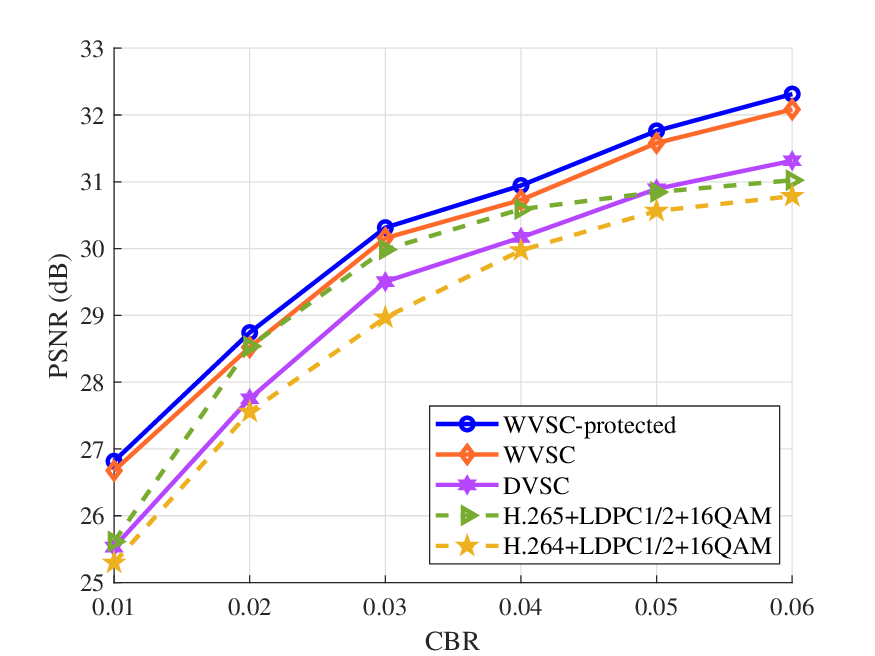}}
	\subfigure[MS-SSIM for the reconstructed videos.]{
		\includegraphics[width=0.32\linewidth]{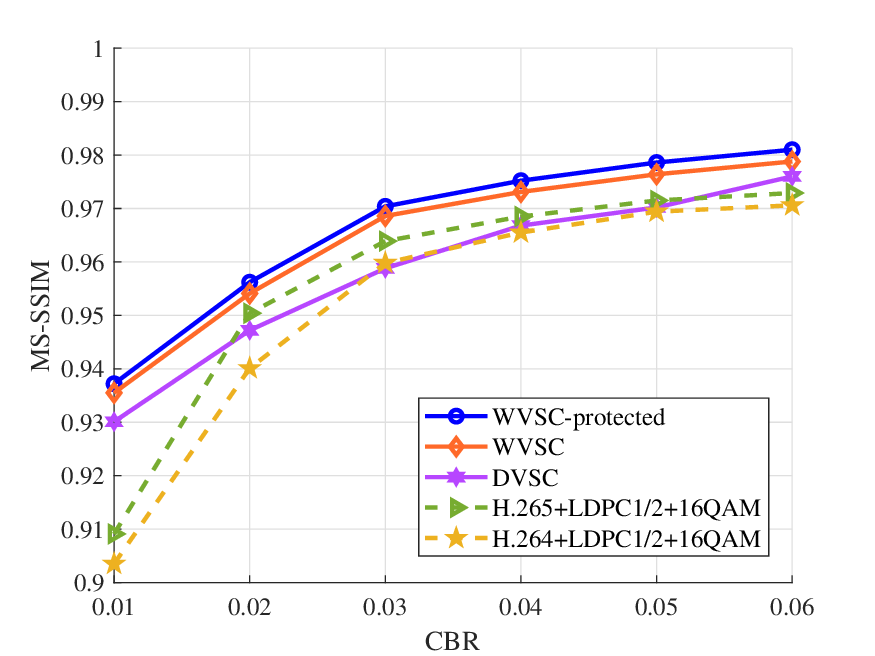}}
	\subfigure[LPIPS for the reconstructed videos.]{
		\includegraphics[width=0.32\linewidth]{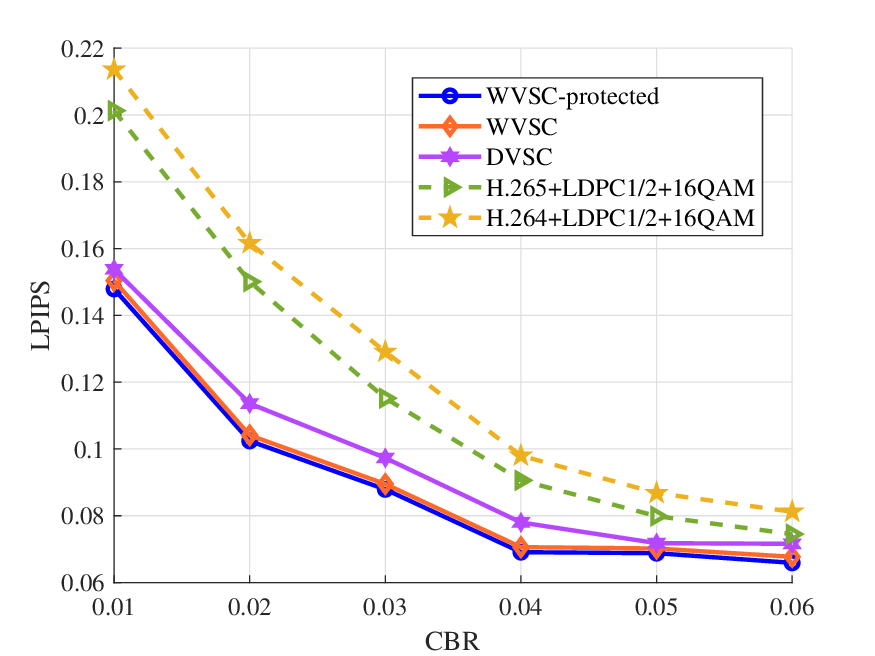}}
	\caption{Quality of the reconstructed images versus the CBRs under Rayleigh fading channels (SNR = 12 dB).}
	\label{fig_6}
\end{figure*}

\subsubsection{Model Deployment Details}
The network deployment of WVSC utilizes the Swin-Transformer \cite{swin} backbone as the semantic codec with $\{N_1,N_2,N_3,N_4\}=\{2,2,2,2\}$ Transformer blocks. Various signal-to-noise ratios (SNRs) are randomly sampled for each image batch during training stage to acquire SNR-adaptive results in various test conditions. For model training, we use variable learning rate, which decreases step-by-step from 1e-4 to 2e-5. The batchsize is set as 1. The default GoP size $N$ is 10. The whole framework is optimized with Adam \cite{Adam} algorithm. All the experiments of WVSC and other DL-based benchmarks are conducted in RTX3090 GPUs with Pytorch2.0.0.

\subsubsection{Comparison Benchmarks}
In the experiments, several benchmarks are given as below

$\textbf{DVSC}$: The DL-empowered deep video transmission framework \cite{dvsc} with SNR-adaptive channel coder and semantic repairment at the receiving end.


$\textbf{H.264+LDPC+QAM}$: The separated source and channel coding scheme with H.264 \cite{264} video codec as source coding with GoP size 32 and 5G Low-density Parity-check (LDPC) \cite{sionna} as channel coding, along with the quadrature amplitude modulation (QAM).

$\textbf{H.265+LDPC+QAM}$: The separated source and channel coding scheme with H.265 \cite{265} video codec as source coding with GoP size 32 and 5G LDPC as channel coding, along with the QAM.

Note that DVSC is the DL-based pixel-level point-to-point wireless transmission scheme, which corresponds to Fig. 1(a). H.265/H.264+LDPC+QAM are SSCC schemes which H.265/H.264 are employed by ffmpeg-python with low delay pattern while 5G LDPC is implemented by \cite{sionna} with code length 4096.
\subsubsection{Evaluation Metrics}

We leverage the widely used pixel-wise metric peak signal-to-noise ratio (PSNR), perceptual-level multi-scale structural similarity (MS-SSIM) \cite{ssim} and learned perceptual image patch similarity (LPIPS) \cite{lpips} as measurements for the reconstructed image quality.  

\subsection{Results Analysis}

\subsubsection{Performance for Different SNRs}
We first evaluate the anti-noise performances of WVSC under Rayleigh fading channels with a specific channel bandwidth ratio (CBR), where $CBR=\frac{L+(N-1)\times L_1}{N\times H\times W\times 3} $. We set it as 0.04. As shown in Fig. \ref{fig_5}(a), it is clearly to observe that WVSC outperforms all other benchmarks with both WVSC and WVSC-protected. Here WVSC and WVSC-protected refer to the 1:1 and 4:1 rate allocation ratio between the I frame and each P frame, respectively. Compared to other DL-based schemes, WVSC outperforms DVSC for about 1 dB in terms of PSNR. This trend verifies that semantic-level wireless video transmission structure seems to be more robust to the channel fading and noise. For the traditional separated coding schemes, two channel coding and modulation parameter pairs are presented. In detail, 2/3 code rate LDPC with 4QAM and 1/2 code rate LDPC with 16QAM are set, which reflect corresponding anti-noise performances under different channel interference levels. Compared to traditional schemes, WVSC provides much more performance gain and stability since traditional schemes would be confronted with serious cliff effect with harsh channel conditions. As shown in Fig. \ref{fig_5}(b) and Fig. \ref{fig_5}(c), the DL-based schemes achieves better reconstruction results in terms of MS-SSIM and LPIPS compared to traditional SSCC schemes, which means that the satisfying visual perception quality is ensured through extracting semantics inside the original video signals. Compared to DVSC, WVSC preserves even much more high frequency image details. These results demonstrate the effectiveness of WVSC for video transmission under Rayleigh fading channels with various noise intensities.

\subsubsection{Performance for Different CBRs}
Then we evaluate the bandwidth compression performances of WVSC under Rayleigh fading channels with SNR = 12 dB. As shown in Fig. \ref{fig_6}(a), WVSC achieves significant performance gain compared to other schemes. Compared to the DVSC, the performance gap increases as CBR increases. Such insight mainly lies in the deployment of video coding in semantic level. Since semantic coding has great potential of source compression, based on the extracted semantics, the video coding enables much more effective data compression performances compared to directly conduct on video signals. In this way, higher CBRs allow the video coding part to retain more semantic information in the form of frame residuals. As the reference frames stay relatively steady as CBR values change, the increased allocation of extra bandwidth resources for transmitting residuals greatly promotes the reconstruction of each semantic frame. For the visual perceptual-level indexes in Fig. \ref{fig_6}(b) and Fig. \ref{fig_6}(c), WVSC retains relatively satisfying visual quality for a wide range of CBRs compared to other schemes. 

%

\begin{figure}[htbp]
	\centering
	\includegraphics[width=3.4in]{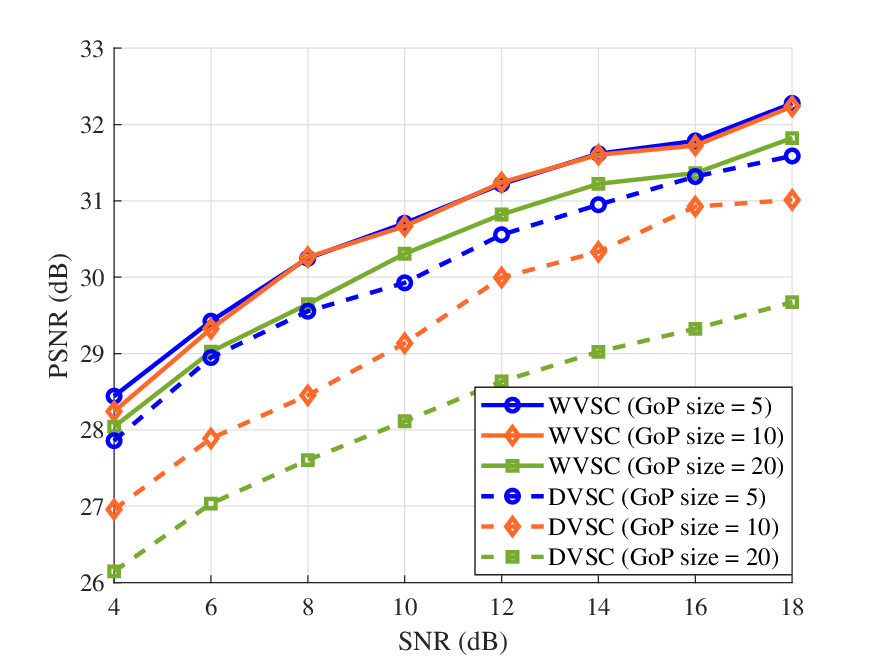}
	\caption{Performance of different GoP sizes. (all trained with GoP size 5)}
	\label{fig_8}
\end{figure}

\subsubsection{Sensitive Analysis of different GoP Sizes}
When utilizing the first frame in a GoP for transmission and video coding, a performance degradation could emerge when the GoP size becomes large. As shown in Fig. \ref{fig_8}, the pretrained framework with GoP size 5 is directly employed to be tested with different GoP sizes. For WVSC, with the multi-frame compensation module, transmitting only the reference frame is enough to achieve satisfying transmission performance, as GoP size 5 and 10 do. With a large GoP size as 20, the overall performance drops but still retains relatively satisfying. With previous reconstructed frames as supplement for compensating motion information, the I frame can be compensated into each P frame. Compared to the DVSC, the performances fluctuate with different GoP sizes. It is due to the intrinsic defect of pixel-level wireless video transmission, which depends much on the accurate motion vectors. With large GoP sizes, DVSC trained with GoP size 5 is unable to match the motion differences between the later adjacent frames. To conclude, when specifically trained with a larger GoP size, we believe WVSC could retain a satisfying performance.

\end{document}